\documentclass[letterpaper, 10 pt, conference]{IEEEtran}
\IEEEoverridecommandlockouts

\usepackage{graphics}
\usepackage{amsmath}
\usepackage{amssymb}
\usepackage{epsfig,wrapfig}
\usepackage{graphicx}
\usepackage{amsmath}
\usepackage{epsfig,wrapfig}
\usepackage{color}
\usepackage{cite}

\title{Comment on `Detecting Topology Variations in Networks of Linear Dynamical Systems'}
\author{Sandip Roy and Mengran Xue
\thanks{The authors acknowledge support from United States National Science Foundation grants 1545104 and 1635184.}
}

\begin{document}
\maketitle

\begin{abstract}
Conditions for the detectability of topology variations in dynamical networks are developed in \cite{tesi}.  Here, an example is presented which illustrates an error in the network-theoretic conditions for detectability developed in \cite{tesi}. \newline \newline
\end{abstract}

The article \cite{tesi} presents conditions under which topology variations in a network of homogeneous linear subsystems can be detected, using measurements of the network's natural response.
The conditions are developed by first characterizing discernibility of the natural responses of a nominal and modified linear system for different initial states (Lemma 1, Corollary 1, and Proposition 1), and then applying this result to the dynamical-network model of interest (Proposition 2 and following results).  The study aims to distill detectability of topology variations into a condition phrased entirely in terms of the network's topology, specifically the spectrum of the Laplacian matrix associated with the network's graph. 

The following example demonstrates an error in the topological results developed in \cite{tesi} (Proposition 2 and following results), and illustrates that the detection of topology variations cannot always be distilled to a condition only on the network's topology.
Per the notation in \cite{tesi}, we consider an example with the following parameters: $A=\begin{bmatrix} 7 & 0 & 0 \\ 0 & 0 & 1 
\\ 1 & 0 & 1 \end{bmatrix}$, $B=\begin{bmatrix} 1 & 1 & -1 \\ 
0 & -1 & 1 \\ 0 & 0 & 0 \end{bmatrix}$, 
$L=\begin{bmatrix} 2 & -1 & -1 & 0 \\ -1 & 2 & -1 & 0 \\ 
-1 & -1 & 3 & -1 \\ 0 & 0 & -1 & 1 \end{bmatrix}$, and 
$\overline{L}=\begin{bmatrix} 1 & -1 & 0 & 0 \\ -1 & 2 & -1 & 0 \\
0 & -1 & 2 & -1 \\ 0 & 0 & -1 & 1 \end{bmatrix}$. We notice that the pair $(A,B)$ is controllable, and also that the Laplacian matrices $L$ and $\overline{L}$ correspond to networks which differ by a single link.  The eigenvalues of $L$ are $\alpha=(0,1,3,4)$, while the eigenvalues of $\overline{L}$ are $\alpha=(0,.59,2,3.4)$.  The two Laplacian matrices thus have only one eigenvalue in common, at $\alpha=0$; the right eigenvectors of the two matrices associated with this eigenvalue are also identical, specifically the vector with all unity entries (${\bf 1}$).  From Proposition 2 and the following development, the non-null indiscernible states of the network model should be a three-dimensional space, corresponding to the synchronized states of the model.  Indeed, the transition matrices $\Phi=I_4 \otimes A-L \otimes B$ and $\overline{\Phi}=I_4 \otimes A -\overline{L} \otimes B$ are seen to have common eigenvalues at $(0,1,7)$ 
whose corresponding eigenvectors are identical, and specify the synchronous manifold.  However, the matrices $\Phi$ and $\overline{\Phi}$ also share an eigenvalue at $1$ whose algebraic multiplicity is $4$.  Further, any vector of the form
$\begin{bmatrix} a \\ b \\ c \\ d \end{bmatrix} \otimes \begin{bmatrix} 0 \\ 1 \\ 1 \end{bmatrix}$ is seen to be an eigenvector of both $\Phi$ and $\overline{\Phi}$ associated with the eigenvalue $1$.  Thus, the non-null indiscernible states form a six-dimensional space, consisting of the synchronous states as well as states of the form $\begin{bmatrix} a \\ b \\ c \\ d \end{bmatrix} \otimes \begin{bmatrix} 0 \\ 1 \\ 1 \end{bmatrix}$.
This disagrees with Proposition 2, Theorem 1, and the ensuing discussion in \cite{tesi}.

The error indicated in the example above arises from
Equation 21 in \cite{tesi}, which claims that the eigenvectors (and generalized eigenvectors) of the transition matrix $\Phi$ are always Kronecker products of the eigenvectors of $L$ and of $A-\alpha B$, where $\alpha \in spec(L)$.  However, this is only necessarily true in the case where the eigenvalues of $A-\alpha_iB$ corresponding to different $\alpha_i \in spec(L)$ are mutually distinct.  Otherwise, if different matrices $A-\alpha_iB$ share eigenvalues, the eigenvectors of $\Phi$ may be linear combinations of such Kronecker-product vectors.  In the example above, the matrix $A-\alpha B$ equals $\begin{bmatrix} 7-\alpha & -\alpha & \alpha \\ 0 & \alpha & 1-\alpha \\ 1 & 0 & 1 \end{bmatrix}$. Thus, the matrix $A-\alpha B$ is seen to have an eigenvalue at $1$ with corresponding right eigenvector $\begin{bmatrix} 0 \\ 1 \\ 1 \end{bmatrix}$, for any complex $\alpha$.  Thus, we immediately recover that $\Phi$ has an eigenvalue at $1$ with multiplicity equal to the number of nodes, and further any vector of the form  $\begin{bmatrix} a \\ b \\ c \\ d \end{bmatrix} \otimes \begin{bmatrix} 0 \\ 1 \\ 1 \end{bmatrix}$ is a right eigenvector associated with the eigenvalue at $1$.  By the same argument, $\overline{\Phi}$ also has the same eigenvalue-eigenvector pairs, and the additional indiscernible states are clarified.

The example suggests that the indiscernible states cannot be determined only based on the topology of the network, since in this case the repeated eigenvalue at $1$ and corresponding eigenspace are present, entirely independently of the network topology.  Thus, any type of topology variation -- including link and node disconnection -- would be indiscernible for some initial states outside the synchronous manifold.

The error in the development may be corrected by adding the technical requirement that the eigenvalues of $A-\alpha_i B$ corresponding to different $\alpha_i$ be distinct.  Alternately, a more complete treatment can perhaps be obtained by either 
pursuing a full eigenvector analysis of the dynamical-network model
(see \cite{dyn1,dyn2}), or by exploiting the concept of a network-invariant mode \cite{netin}.  We also note that the subtlety in the eigenvector analysis of the dynamical-network model discussed here has led to errors in the controllability analysis of the model (e.g. \cite{control1}).

\end{document}